\documentclass[11pt]{article}
\newtheorem{Def}{Def.}[section]
\newtheorem{Thm}[Def]{Theorem}

\newtheorem{Lemma}[Def]{Lemma}

\newcommand{\Proof}{{\em{Proof: }}}
\newcommand{\QED}{\ \hfill $\FBox$ \\[1em]}

\newcommand{\Tr}{\mbox{Tr\/}}

\newcommand{\Pdd}{\mbox{$\partial$ \hspace{-1.2 em} $/$}}

\newcommand{\Sl}{\mbox{$\prec \!\!$ \nolinebreak}}
\newcommand{\Sr}{\mbox{\nolinebreak $\succ$}}
\newcommand{\bra}{\mbox{$< \!\!$ \nolinebreak}}
\newcommand{\ket}{\mbox{\nolinebreak $>$}}

\newcommand{\spc}{\;\;\;\;\;\;\;\;\;\;}

\newcommand{\C}{\mbox{\rm I \hspace{-1.25 em} {\bf C}}}
\newcommand{\R}{\mbox{\rm I \hspace{-.8 em} R}}
\newcommand{\1}{\mbox{\rm 1 \hspace{-1.05 em} 1}}

\newcommand{\sR}{\mbox{\rm \scriptsize I \hspace{-.8 em} R}}

\newcommand{\FBox}{\rule{2mm}{2.25mm}}

\title{Local $U(2,2)$ Symmetry in Relativistic Quantum 
Mechanics\thanks{Supported by the Deutsche Forschungsgemeinschaft, Bonn.}}
\author{Felix Finster\\ Harvard University, Department of Mathematics}
\date{March 1997 / May 1998}

\begin{document}
\maketitle
\begin{abstract}
Local gauge freedom in relativistic quantum mechanics is derived from a
measurement principle for space and time. For the Dirac equation, one obtains
local $U(2,2)$ gauge transformations acting on the spinor index
of the wave functions. This local $U(2,2)$ symmetry allows a unified description
of electrodynamics and general relativity as a classical gauge theory.
\end{abstract}

\section{Connection between Local Gauge Freedom and Position Measurements}
In \cite{F1}, it was suggested to link the physical gauge principle with
quantum mechanical measurements of the position variable.
In the present paper, we will extend this concept to relativistic quantum
mechanics and apply it to the Dirac equation. For Dirac spinors, we 
obtain local $U(2,2)$ gauge freedom. Our main result is that this 
$U(2,2)$ symmetry allows a natural description of both 
electrodynamics and general relativity as a classical gauge theory. 
This is shown by deriving a $U(2,2)$ spin connection from the Dirac 
operator and analyzing the geometry of this connection. Although we 
develop the subject from a particular point of view, this paper can 
be used as an introduction to the Dirac theory in curved space-time.

In contrast to \cite{F1}, where the point of interest is the measure 
theoretic derivation of local gauge transformations, we will here 
concentrate on the differential geometry of the Dirac operator. In 
order to keep measure theory out of this paper, we will use a bra/ket 
notation in position space. This allows us to explain the basic ideas 
and results of \cite{F1} in a simple, non-technical way which will be 
sufficient for the purpose of this paper. Nevertheless one should keep 
in mind that the bra/ket symbols and the $\delta$-normalizations are 
only a formal notation; the mathematical justification for this 
formalism is given in \cite{F1}.

We begin with the example of a scalar particle in nonrelativistic quantum
mechanics. The particle is described by a wave function $\Psi(\vec{x})$, which
is a vector of the Hilbert space $H=L^2(\R^3)$. The physical observables
correspond to
self-adjoint operators on $H$. The position operators $\vec{X}$, for example,
are given as multiplication operators with the coordinate functions,
$X^i \::\: \Psi(\vec{x}) \rightarrow x^i \: \Psi(\vec{x})$, $i=1,\ldots,3$.
Our definition of the Hilbert space as a space of functions
in the position variable was only a matter of convenience; e.g.\ we could
just as well have introduced $H$ as functions in momentum space.
Therefore it seems reasonable to forget about the fact that $H$ is a function
space and consider it merely as an abstract Hilbert space.
After this generalization, which is often implicitly assumed in quantum mechanics,
we must construct the representation of $H$ as wave functions.
For this purpose, we choose an ``eigenvector basis'' $|\vec{x} \ket$ of the
position operators,
\begin{equation}
\label{1_1}
X^i \:|\vec{x} \ket \;=\; x^i \: |\vec{x} \ket \;\;\;,\spc
	\bra \vec{x} \:|\: \vec{y} \ket \;=\; \delta^3(\vec{x} - \vec{y}) \spc ,
\end{equation}
and define the wave function of a vector $\Psi \in H$
by $\Psi(\vec{x}) = \bra \vec{x} \:|\: \Psi \ket$.

This ``position representation'' of the Hilbert space is quite elementary and
is currently used in physics. We point out that it is not unique, because
the ``eigenvectors'' $|\vec{x} \ket$ are only determined up to a phase. 
Namely, we can transform $|\vec{x} \ket$ according to
\begin{equation}
\label{1_2}
|\vec{x} \ket \;\rightarrow\; e^{i e \Lambda(\vec{x})} \: |\vec{x} \ket
\end{equation}
with a real function $\Lambda(\vec{x})$. This corresponds to a local phase
transformation
\begin{equation}
\label{1_3}
\Psi(\vec{x}) \;\rightarrow\; e^{-i e \Lambda(\vec{x})} \: \Psi(\vec{x})
\end{equation}
of the wave functions.
The arbitrariness of the local phase of the wave functions can also be understood
directly from the fact that the wave function itself is not observable,
only its absolute square $|\Psi|^2$ has a physical interpretation as probability
density.
It is important for the following that the local phase transformations
(\ref{1_2}),(\ref{1_3}) can be interpreted as $U(1)$ gauge transformations.
To see this, we consider the canonical momentum operator $\vec{\pi}=-i \vec{\nabla}
- e \vec{A}$ with magnetic vector potential $\vec{A}$.
Under the transformation (\ref{1_2}), the canonical momentum behaves like
\[ \vec{\pi} \;\rightarrow\; e^{-ie \Lambda(\vec{x})} \:\vec{\pi}\:
	e^{ie \Lambda(\vec{x})} \;=\; -i \vec{\nabla}
	- e (\vec{A} - \vec{\nabla} \Lambda) \spc , \]
which corresponds to the gauge transformation $\vec{A} \rightarrow \vec{A} -
\vec{\nabla} \Lambda$ of the vector potential. 

In this way, we have explained the local $U(1)$ gauge symmetry of the magnetic
field from the fact that space is a quantum mechanical observable.
Unfortunately, this natural explanation of local $U(1)$ gauge freedom only 
works for scalar particles. In the general case with spin, the wave functions
have several components, $\Psi \in H = L^2(\R^3)^m$ (i.e.\ $m=2s+1$ for particles
with spin $s$). We again consider $H$ as an abstract Hilbert space.
In order to construct the representation of
a vector $\Psi \in H$ as a function, we choose an ``orthonormal basis''
$|\vec{x} \alpha \ket$, $\vec{x} \in \R^3$, $\alpha=1,\ldots,m$ of the position
operators,
\begin{equation}
\label{1_4b}
X^i \:|\vec{x} \alpha \ket \;=\; x^i \:|\vec{x} \alpha \ket \;\;\;\;,\spc
	\bra \vec{x} \alpha \:|\: \vec{y} \beta \ket \;=\; \delta_{\alpha \beta} \:
	\delta^3(\vec{x} - \vec{y}) \spc ,
\end{equation}
and define the wave function by $\Psi^\alpha(\vec{x}) = \bra \vec{x} \alpha \:|\:
\Psi \ket$. The basis $|\vec{x} \alpha \ket$ is only unique up to local unitary
transformations,
\begin{equation}
\label{1_4a}
|\vec{x} \alpha \ket \;\rightarrow\; \sum_{\beta=1}^m (U^{-1})^\alpha_\beta(\vec{x}) \:
	|\vec{x} \beta \ket \spc {\mbox{with}} \spc U(\vec{x}) \in U(m) \spc,
\end{equation}
as is verified by substituting into (\ref{1_4b}). Thus the wave functions
can be transformed according to
\begin{equation}
\label{1_4}
\Psi(\vec{x}) \;\rightarrow\; U(\vec{x}) \: \Psi(\vec{x}) \spc .
\end{equation}

It would be nice if we could again identify the local
$U(m)$ transformations (\ref{1_4a}),(\ref{1_4}) with physical gauge
transformations. Then the local gauge principle would no longer be an 
a-priori principle in physics. It would be a consequence of a quantum 
mechanical ``measurement principle'', namely the description of space 
with observables $X^i$ on an abstract Hilbert space.
Notice that the local $U(m)$ transformations (\ref{1_4}) 
are a generalization of the phase transformations (\ref{1_3}). Our 
idea is that the additional degrees of freedom of the larger gauge 
group in (\ref{1_4}) might make it possible to describe additional 
interactions (like gravitation or the weak and strong forces). Notice 
that, as a great advantage of our procedure, the local gauge group 
could no longer be chosen arbitrarily; it would be determined by the
configuration of the spinors. Thus we could hope to get into the 
position to explain the gauge groups observed in physics.
At the moment, however, it is not clear if the interpretation of 
(\ref{1_4a}),(\ref{1_4}) as gauge transformations really makes 
physical sense. We will in the following simply assume it as a postulate and
want to study its further consequences.

In view of our later generalization to curved space (and curved space-time), it is
convenient to introduce the spectral measure $dE_{\vec{x}}$ of the position
operators:
We form the ``projectors'' $E_{\vec{x}}$ on the ``eigenspaces'' of $\vec{X}$,
\begin{equation}
\label{2_19}
E_{\vec{x}} \;=\; \sum_{\alpha=1}^m |\vec{x} \alpha \ket \bra \vec{x} \alpha | \spc .
\end{equation}
The $E_{\vec{x}}$ do not depend on the choice of
$|\vec{x} \alpha \ket$; we can characterize them by the operator relations
\begin{equation}
\label{2_20}
X^i \: E_{\vec{x}} \;=\; x^i \: E_{\vec{x}} \;\;\;,\spc
	E_{\vec{x}} \: E_{\vec{y}} \;=\; \delta^3(\vec{x}-\vec{y}) \: E_{\vec{x}}
	\spc .
\end{equation}
The $\delta^3$-normalization in (\ref{2_20}) leads to difficulties as soon as curved
coordinate systems are considered. To avoid these problems, it is useful to
combine $E_{\vec{x}}$ with the integration measure by introducing the spectral
measure $dE_{\vec{x}} := E_{\vec{x}} \: d\vec{x}$.
Integrating over the spectral measure
yields an operator on the Hilbert space. For example, we have
\begin{equation}
\label{2_21}
X^i \;=\; \int_{\sR^3} x^i \; dE_{\vec{x}} \;\;\;,\spc
	\chi_V \;=\; \int_V dE_x \spc ,
\end{equation}
where $\chi_V$ is the multiplication operator with the characteristic function
(i.e.\ $(\chi_V \Psi)(x)$ is equal to $\Psi(x)$ if $x \in V$ and vanishes otherwise).
These relations can be verified directly with the help of
(\ref{1_4b}) and (\ref{2_19}).
Actually, the spectral measure is completely characterized by (\ref{2_21}).
In the mathematical paper \cite{F1}, spectral measures on a manifold are used as
the starting point. In this more general approach, the number of components
of the wave functions $m=m(x)$ may vary in space. We call it the
{\em{spin dimension}}.

\section{Generalization to Dirac Spinors, $U(2,2)$ Gauge Symmetry}
If our explanation of local gauge freedom shall have general significance,
it must be possible to extend it to the relativistic setting.
We start the analysis with Dirac wave functions in Minkowski space\footnote{We
mention for completeness that {\em{Minkowski space}} is a four-dimensional real vector
space with a scalar product of signature $(+---)$. A coordinate system where this
scalar product coincides with the Minkowski metric $\eta_{ij}$ is called
{\em{reference frame}}. We will mainly work in a fixed reference frame and can then
identify Minkowski space with $\R^4$.}.
Since we do not yet want to introduce Dirac matrices and the Dirac operator,
we consider the wave functions $\Psi, \Phi, \ldots$ just as $4$-component
functions. We define a scalar product on the spinors,
\begin{equation}
\label{2_a}
\Sl \Psi, \Phi \Sr(x) \;=\; \sum_{\alpha=1}^4 s_\alpha \:
	(\Psi^\alpha(x))^* \: \Phi^\alpha(x) \spc {\mbox{with}} \spc s_1=s_2=1,\;\;
	s_3=s_4=-1 \;\;\; ,
\end{equation}
which is called {\em{spin scalar product}}. It is indefinite of signature $(2,2)$.
Later, after the introduction of the Dirac matrices in Dirac representation,
the spin scalar product will coincide with $\overline{\Psi} \Phi$, where
$\overline{\Psi}=\Psi^* \gamma^0$ is the adjoint spinor.
Our definition without referring to Dirac matrices
will clarify the relation between coordinate and gauge transformations
in section \ref{sec3}.

The basic question is how we want to describe space-time. At the moment, position
and time are merely parameters of the wave functions. This is the usual description in
relativistic quantum mechanics. In the previous section, however, it 
was essential for the explanation of local gauge transformations that 
space corresponds to some operators on the wave functions.
In order to generalize this ``measurement principle'' to the 
relativistic setting, we introduce ``observables'' for space and time as
multiplication operators with the coordinate functions,
\begin{equation}
\label{2_3}
X^i \;:\; \Psi(x) \:\rightarrow\: x^i \: \Psi(x) \;\;\;,\spc i=0,\ldots,3 \spc .
\end{equation}
We point out that these operators, especially the time operator $X^0$, are
commonly not used in relativistic quantum mechanics. Nevertheless our 
description of Minkowski space makes sense, as we will see in the 
following.

The next step is the introduction of a scalar product on the wave functions.
Usually one considers the positive product
\begin{equation}
\label{2_1}
(\Psi \:|\: \Phi) \;=\; \int_{\sR^3} \Psi^*(t,\vec{x}) \: \Phi(t,\vec{x}) \;
	d\vec{x} \spc ,
\end{equation}
where the spinors are integrated over a space-like hypersurface $t={\mbox{const}}$
(we avoid the notation $\Psi^* \Phi = \overline{\Psi} \gamma^0 \Phi$ with
the adjoint spinor because we have no Dirac matrices at the moment).
The integrand of $(\Psi \:|\: \Psi)$ can be interpreted as the probability density
of the particle. For solutions of the Dirac equation, current conservation implies
that (\ref{2_1}) is independent of $t$.
As an apparent problem of this scalar product, time measurements do not make sense,
because expectations $(\Psi \:|\: X^0 \:|\: \Psi) = t \:(\Psi \:|\: \Psi)$
depend on the choice of the hypersurface.
The only way out is to introduce a different scalar product
where the spinors are also integrated over the time variable, namely
\begin{equation}
\label{2_2}
\bra \Psi \:|\: \Phi \ket \;=\; \int_{\sR^4} \Sl \Psi \:|\: \Phi \Sr \: d^4x \spc .
\end{equation}
In contrast to (\ref{2_1}), the scalar product (\ref{2_2}) has no 
immediate physical interpretation. Nevertheless it causes no problem 
to consider (\ref{2_2}) as the fundamental scalar product on the 
spinors, for the following reason:
The scalar product does not enter in the physical equations (i.e.\ the
Dirac equation and the classical field equations), we only need it for the
probabilistic interpretation of the wave functions. For this final
interpretation, we can just introduce (\ref{2_1}) by an additional mathematical
construction (see equation (\ref{SP})).

We choose (\ref{2_2}) as the fundamental scalar product on the wave functions.
We denote the corresponding function space
by $(H, \bra .|. \ket)$ and consider it together with the position/time operators
(\ref{2_3}) as an abstract scalar product space. In order to represent the vectors of $H$
as wave functions, we choose an ``eigenvector basis'' $|x \alpha \ket$, $x \in \R^4$,
$\alpha=1,\ldots,4$ of the position/time operators,
\begin{equation}
\label{2_4}
X^i \:|x \alpha \ket \;=\; x^i \:|x \alpha \ket \;\;\;,\spc
\bra x \alpha \:|\: y \beta \ket \;=\; s_\alpha \:\delta_{\alpha \beta}\:
	\delta^4(x-y) \spc ,
\end{equation}
and define for $\Psi \in H$ the corresponding wave function by
$\Psi^\alpha(x) = \bra x \alpha \:|\: \Psi \ket$.
This is a relativistic analogue of (\ref{1_4b}).
The additional factors $s_\alpha=\pm1$ occur as a consequence of the indefinite
spin scalar product. The arbitrariness of the choice of $|x \alpha \ket$ leads to
the transformations
\begin{eqnarray}
|x \alpha \ket &\rightarrow& \sum_{\beta=1}^4 (U^{-1})^\alpha_\beta(x) \:|x \beta \ket
	\nonumber \\
\label{2_11}
\Psi(x) &\rightarrow& U(x) \: \Psi(x)
	\spc {\mbox{with}} \spc U(x) \in U(2,2) \spc ,
\end{eqnarray}
which we interpret as local $U(2,2)$ gauge transformations.
According to this interpretation, we call $|x \alpha \ket$ a {\em{gauge}}.
Notice that the gauge group is non-compact, which is directly related to the
indefiniteness of the spin scalar product.
Similar to the nonrelativistic case, the spectral measure is given by
\[ dE_x \;=\; \sum_{\beta=1}^4 s_\beta \: |x \beta \ket \bra x \beta| \; d^4x \spc . \]
We say that the spin dimension is $(2,2)$.

We remark that the indefiniteness of the scalar product $\bra .|. \ket$ leads to
mathematical problems in the rigorous derivation of gauge transformations. A first, but
not fully convincing attempt towards a satisfying mathematical formulation was made
in \cite{F0} (which also contains a preliminary version of \cite{F1}).

\section{The Free Dirac Operator, Lorentzian Transformations}
\label{sec3}
Having introduced the Dirac wave functions in Minkowski space, we can now define the
free Dirac operator: We fix the reference frame, choose a special 
gauge, and introduce the partial differential
operator $i \Pdd := i \gamma^j \partial_j$ on the wave functions, where $\gamma^j$
denote the usual $(4 \times 4)$-matrices in the Dirac representation acting on
the spinor index,
\[ \gamma^0 \;=\; \left( \begin{array}{cc} \1 & 0 \\ 0 & -\1 
	\end{array} \right) \;\;\;,\spc \gamma^i \;=\;
	\left( \begin{array}{cc} 0 & \sigma^i \\ -\sigma^i & 0
	\end{array} \right) \;\;,\;\;\; i=1,2,3 \]
($\sigma^1, \sigma^2, \sigma^3$ are the Pauli matrices).
Now we forget about the fact that the free Dirac operator was defined in a
special gauge and a special reference frame and consider it as an operator on $H$.
We denote this operator by $G$ and write the free Dirac equation in the coordinate
and gauge invariant form $(G-m) \Psi = 0$.

The Dirac operator is Hermitian (with respect to the scalar product
$\bra .|. \ket$). This is most conveniently verified in the original
gauge and reference frame where $G=i \Pdd$: For simplicity, we rewrite the
factors $s_\alpha$ in (\ref{2_a}) with the matrix $\gamma^0$ and obtain
\begin{eqnarray*}
\bra G \Psi \:|\: \Phi \ket &=& \int_{\sR^4} \Sl G \Psi \:|\: \Phi \Sr_{|x} \: d^4x
\;=\; \int_{\sR^4} (i \gamma^j \partial_j \Psi)(x)^* \:\gamma^0\: \Phi(x) \: d^4x \\
&=& -i \int_{\sR^4} (\partial_j \Psi)^* \:(\gamma^j)^* \:\gamma^0\: \Phi \:d^4x
\;=\; i \int_{\sR^4} \Psi^* \:\gamma^0\: (\gamma^0 \:(\gamma^j)^*\: \gamma^0)
	\: \partial_j \Phi \\
&=& i \int_{\sR^4} \Psi^* \:\gamma^0\: \gamma^j \partial_j \Phi
\;=\; \int_{\sR^4} \Sl \Psi \:|\: G \Phi \Sr_{|x} \: d^4x \;=\; \bra \Psi \:|\:
	G \Phi \ket \spc .
\end{eqnarray*}
Note that the $\gamma$-matrices are not Hermitian (more precisely, $(\gamma^0)^*
=\gamma^0$ and $(\gamma^j)^*=-\gamma^j$ for $j=1,2,3$), but they are self-adjoint
with respect to the spin scalar product, $\Sl \gamma^j \Psi \:|\: \Phi \Sr
= \Sl \Psi \:|\: \gamma^j \Phi \Sr$.

We point out that $G$ is as operator on $H$ a coordinate and
gauge independent mathematical object (although its definition might depend on
the special gauge and reference frame chosen at the beginning).
Both coordinate and gauge transformations
are merely passive transformations and lead to equivalent representations of $G$.
Especially, transformations of the space-time coordinates
and of the spinors are independent of each other. This is a major
difference to the usual description of the Dirac equation.
Therefore we will now discuss coordinate and gauge transformations
and explain what ``Lorentzian invariance'' of the free Dirac operator exactly
means in our setting.

We first look at the Dirac operator in a general gauge, but still in the original
reference frame. The gauge transformation (\ref{2_11}) leads to
\begin{eqnarray*}
i \Pdd \;\longrightarrow\; G &=& U \:i \Pdd\: U^{-1} \;=\; i G^j
	\frac{\partial}{\partial x^j} + B \spc {\mbox{with}} \\
G^j(x) &=& U(x) \:\gamma^j\: U(x)^{-1}\;\;\;,\spc
	B(x) \;=\; i U(x) \:\gamma^j\: (\partial_j U(x)^{-1}) \spc .
\end{eqnarray*}
As explained before, all these operators are equivalent; of course
it is most convenient to work in the original gauge with $G=i \Pdd$.

Next we look at a Lorentzian transformation. The transformation of the space-time
variables is described by a coordinate transformation $x^j \rightarrow \Lambda^j_k x^k$,
where $\Lambda$ is an isometry of Minkowski space. The Dirac operator transforms like
\begin{equation}
\label{3_12}
i \Pdd \:\longrightarrow\: G \:=\: i \gamma^j \Lambda^k_j\: \partial_k \spc .
\end{equation}
Thus in the new reference frame, the Dirac operator no longer has the original form.
This might seem to contradict Lorentzian invariance, but we must keep in mind
that all operators related to each other by gauge transformations are equivalent.
Thus we must look for a gauge transformation such that the Dirac
operator in the new reference frame again coincides with $i \Pdd$.
As is shown in standard textbooks on relativistic quantum mechanics (see e.g.\ \cite{BD}),
there is a $(4 \times 4)$-matrix $U(\Lambda)$ with
\[ U(\Lambda) \:\gamma^j \Lambda^k_j \:U(\Lambda)^{-1} \;=\; \gamma^k \spc . \]
It is important for our purpose that this matrix is unitary with respect
to the spin scalar product, $U(\Lambda) \in U(2,2)$.
Therefore we can perform a gauge transformation with
$U(x) \equiv U(\Lambda)$. In the new gauge we again have $G=i \Pdd$, which shows
Lorentzian invariance of $G$.
To summarize, we describe a Lorentzian transformation in two steps.
The first step is a coordinate, the second a gauge transformation. If both steps
are performed at once, the transformation of the space-time coordinates and
of the spinors are related in the usual way.
Notice, however, that the second step is only a matter of convenience;
we could just as well work in the new reference frame in any other gauge.
This splitting of the space-time and spinor transformation becomes possible
because the conditions $G^j=\gamma^j$ give a link between coordinate and gauge
transformations.

We remark that it is not trivial that a spinorial equation can be described with
passive coordinate and gauge transformations. As a counterexample, we consider the
Weyl equation:
Chiral fermions are usually described by two-component spinors $\chi$ satisfying
the equation
\begin{equation}
\label{3_14}
i \sigma^j \frac{\partial}{\partial x^j} \: \chi \;=\; 0
\end{equation}
with Pauli matrices $\sigma^0 = \1$, $\vec{\sigma}$.
This equation is Lorentzian invariant in the sense that there is a one-to-one
correspondence between the solutions of (\ref{3_14}) in different reference frames
(see e.g.\ \cite{BD}). We assume that the Weyl operator $W:=i \sigma^j \partial_j$ were
given as a coordinate and gauge independent object (we will not specify the scalar
product on the wave functions).
After a Lorentzian transformation $x^j \rightarrow \Lambda^j_k x^k$, the Weyl
operator has the form
$W = i \sigma^k \Lambda^j_k \: \partial_j$.
If $W$ was Lorentzian invariant, it should be possible to perform a gauge
transformation such that $W=i \sigma^j \partial_j$. Thus there should be a matrix
$U(\Lambda)$ with
\[ U(\Lambda) \:\sigma^k \Lambda^j_k\: U(\Lambda)^{-1} \;=\; \sigma^j \spc . \]
Taking the trace on both sides, however, yields the equation
$\Lambda^j_0 = \delta^j_0$, which is only satisfied for $\Lambda = \1$.
We conclude that the Weyl operator cannot be described as a Lorentz invariant
operator in our setting (this is not a serious problem because the
Weyl equation can always be obtained as the left- or right-handed component of the
massless Dirac equation).

\section{General Definition of the Dirac Operator}
In the previous section, we introduced the free Dirac operator as an operator on
the wave functions in Minkowski space.
We point out that this definition involved some global assumptions
on space-time. First of all, the reference frames gave global coordinate systems.
Furthermore, we chose a special gauge at the beginning and described Lorentzian
transformations using global gauge transformations (i.e.\ transformations
independent of $x$).
According to the principle of general relativity, however, global conditions do not
make sense. In this section, we replace them by corresponding local conditions,
which will yield the general definition of the Dirac operator.

We start with the observation that the Minkowski metric can be derived from the
Dirac matrices by $\eta^{ij} \:\1 = \frac{1}{2} \: \{G^i,\: G^j\}$.
This allows us to forget about the Minkowski metric and view
Minkowski space simply as a four-dimensional vector space.
For the generalization to curved coordinate systems, we replace this vector
space by a four-dimensional (smooth) manifold $M$. Notice that $M$ has no
Lorentzian structure; we will later recover the Lorentzian metric from the Dirac
operator.

Next we must generalize the scalar product space $(H, \bra.|.\ket)$ and
the position/time operators to the setting of a space-time manifold.
This can be most easily done with the help of spectral measures:
We consider an indefinite scalar product space $(H, \bra.|.\ket)$ and
introduce a spectral measure $(dE_x)_{x \in M}$ on $M$ of spin dimension $(2,2)$.
For any chart $(x^i, U)$, we define corresponding position/time operators by
\begin{equation}
\label{4_1}
X^i \;=\; \int_U x^i \: dE_x \spc .
\end{equation}
We choose an ``eigenvector basis'' $|x \alpha \ket$, $x \in U$,
$\alpha=1,\ldots,4$ of the $X^i$,
\begin{equation}
\label{4_2}
X^i \:|x \alpha \ket \;=\; x^i \:|x \alpha \ket \;\;\;,\spc
	\bra x \alpha \:|\: y \beta \ket \;=\; s_\alpha \: \delta_{\alpha \beta} \:
	\delta^4(x-y) \spc ,
\end{equation}
and represent a vector $\Psi \in H$ as the wave function $\Psi^\alpha(x)=\bra x \alpha
\:|\: \Psi \ket$, $x \in U$.
This construction generalizes (\ref{2_21}),(\ref{2_4}) to curved coordinate systems.
The spectral measure in (\ref{4_1}) allows us to extend the usual changes of charts to the
level of a calculus of operators.

After these preparations, we can introduce the Dirac operator $G$. First of all,
it shall be a Hermitian operator on $H$. In a representation of $H$ as wave
functions, we assume it to be a partial differential operator of first
order, i.e.
\begin{equation}
\label{4_6}
G \;=\; i G^j \frac{\partial}{\partial x^j} + B
\end{equation}
with $(4 \times 4)$-matrices $G^j(x)$,$B(x)$. Since $G$ is (as an operator on $H$)
a coordinate and gauge independent object, the matrices $G^j(x)$ and $B(x)$
have a well-defined behavior under coordinate and gauge transformations.
As additional condition, we demand that the matrices $G^j$ shall coincide
locally with the usual Dirac matrices:
\begin{Def}
\label{def1}
A partial differential operator $G$ on $H$ is called {\bf{Dirac operator}}, if
for any $p \in M$ there is a chart $(x^i, U)$ around $p$ and a gauge $|x \alpha \ket$
such that $G$ has the form (\ref{4_6}) with
\begin{equation}
\label{4_21}
G^j(p) \;=\; \gamma^j \spc .
\end{equation}
\end{Def}
This definition can also be understood in more physical terms: \label{obs}
Let us assume that an observer (e.g.\ a spacecraft passing through the
space-time point $p$) chooses a coordinate system and gauge satisfying
(\ref{4_21}). Since the Dirac matrices at $p$ coincide with those of
the free Dirac operator, the observer at $p$ has the impression that space-time
locally resembles Minkowski space. Thus we expect
that the Dirac operator gives a Lorentz metric
$g_{jk}$, which coincides with the Minkowski metric at $p$, $g_{jk}(p)=\eta_{jk}$.
We can really derive this Lorentz metric from $G$; it is most conveniently
given by
\begin{equation}
\label{4_4}
g^{jk}(x) \:\1 \;=\; \frac{1}{2} \: \{ G^j(x),\: G^k(x) \} \spc .
\end{equation}
In this way, our definition of the Dirac operator incorporates the principle
of general relativity; the coordinate and gauge satisfying (\ref{4_21})
gives a local reference frame.
If the partial derivatives $\partial_j g_{kl}(p)$ of the metric also vanish
(which can be arranged by choosing a so-called ``normal coordinate system''),
the reference frame is a ``local system of inertia'', where the observer feels no
gravitational force at $p$.
It might happen that $\partial_j G^k(p) \neq 0$ although $\partial_j g_{kl}(p)$
vanishes. This corresponds to a ``spin-mixing force'' which is related to our
$U(2,2)$ symmetry. The matrix $B(x)$ can be used for the description of additional
interactions, especially of electromagnetism $G = i \Pdd + e A\!\!\!\slash$.
Note that $B(x)$ was not at all specified in our definition of the
Dirac operator. This will turn out to be too general for physical applications.
We postpone the discussion and specification of $B(x)$ to section \ref{sec6}.

Unfortunately, the equation (\ref{4_2}) for the ``eigenvector basis'' 
is too simple and must be modified. This is only a minor technical point, but
nevertheless we give it in detail:
The $\delta^4$-distribution in (\ref{4_2}) is useful in combination with the integration
measure $d^4x$, because integrations can be easily carried out using the 
relation $\int f(x) \: \delta^4(x) \: d^4x = f(0)$.
On a manifold, however, one works with the invariant measure $d\mu := \sqrt{|g|} \:d^4x$
instead of $d^4x$. In order to compensate the factor $\sqrt{|g|}$ of 
the integration measure in 
the integral over the $\delta$-distribution, we must view the 
combination $|g(y)|^{-\frac{1}{2}} \: \delta^4(x-y)$ as the 
``invariant $\delta$-distribution'' at a point $y$ of the manifold. 
Thus we are led to replace (\ref{4_2}) by
\begin{equation}
\label{4_3}
X^i \:|x \alpha \ket \;=\; x^i \:|x \alpha \ket \;\;\;,\spc
	\bra x \alpha \:|\: y \beta \ket \;=\; s_\alpha \: \delta_{\alpha \beta} \:
	\frac{1}{\sqrt{|g|}} \: \delta^4(x-y) \spc .
\end{equation}
This definition of a gauge is much better than (\ref{4_2}) because
$|x \alpha \ket$ now behaves under coordinate transformations
$x \rightarrow x^\prime(x)$ simply like
$|x \alpha \ket \rightarrow |x^\prime \alpha \ket$.
We again introduce the wave functions by $\Psi^\alpha(x)=\bra x \alpha \:|\: \Psi
\ket$ and obtain a representation of $G$ as the differential operator (\ref{4_6}).
Notice that this indirect definition of a gauge by first introducing (\ref{4_2})
and then replacing it by (\ref{4_3}) was necessary because we had no invariant
measure at the beginning.
We must check that this procedure is consistent:
The transition from (\ref{4_2}) to (\ref{4_3}) is described by the 
transformation
$|x \alpha \ket \rightarrow |g(x)|^{-\frac{1}{4}} \: |x \alpha \ket$.
Consequently, the Dirac operator transforms as
$G \rightarrow |g(x)|^{-\frac{1}{4}} \:G\: |g(x)|^{\frac{1}{4}}$,
and thus $G^j \rightarrow G^j$, $B \rightarrow B + \frac{i}{4} G^j (\partial_j
\log |g|)$. We conclude that the definition of the Dirac operator and the
Lorentz metric (\ref{4_4}) are independent of whether (\ref{4_2}) or (\ref{4_3})
are used.

We will in the following always work with (\ref{4_3}) as the definition of a 
{\em{gauge}}. The freedom in the choice of the gauge again describes local 
$U(2,2)$ transformations of the wave functions
\begin{equation}
	\Psi \;\rightarrow\; U(x) \: \Psi(x) \spc \mbox{with} \spc U(x) \in 
	U(2,2) \spc .
	\label{4_30}
\end{equation}
We define the spin scalar product by (\ref{2_a}). According to 
(\ref{4_3}), the scalar product is obtained by integrating over the 
spin scalar product,
\begin{equation}
	\bra \Psi \:|\: \Phi \ket \;=\; \int_M \Sl \Psi \:|\: \Phi \Sr \; d\mu
	\spc .
	\label{4_00}
\end{equation}
The Dirac equation is again introduced as the operator equation
\begin{equation}
	(G-m) \: \Psi \;=\; 0 \spc .
	\label{4_31}
\end{equation}
On solutions of the Dirac equation, we finally define a 
generalization of the scalar product (\ref{2_1}): We choose a 
space-like hypersurface ${\cal{H}}$ with normal vector field $\nu$ and set
\begin{equation}
(\Psi \:|\: \Phi)_{\cal{H}} \;=\; \int_{\cal{H}} \Sl \Psi \:|\: G^j 
\:|\: \Phi \Sr \:\nu_j \; d\mu_{\cal{H}} \spc ,
	\label{SP}
\end{equation}
where $d\mu_{\cal{H}}$ is the measure on ${\cal{H}}$ induced by the 
Lorentzian metric.

\section{Construction of the Spin Derivative}
We just introduced the Dirac operator on a manifold by
combining the principle of general relativity with our $U(2,2)$ gauge
symmetry. Although this definition was very natural, it is not clear what
it precisely means:
The Dirac operator is characterized by the matrices $G^j(x)$ and $B(x)$ in
(\ref{4_6}). We saw that $G^j(x)$ induces a Lorentzian structure and
that $B(x)$ is composed of potentials which might be identified with
usual gauge potentials. The precise physical interpretation of the 
degrees of freedom of $G^j(x)$,$B(x)$ is not clear, however.
Furthermore, we expect in analogy to the discussion of Lorentzian 
transformations in section \ref{sec3} that coordinate and gauge 
transformations can be linked by imposing some conditions on the Dirac 
matrices. This would imply that the freedom in choosing the coordinate
system is related to the $U(2,2)$ gauge symmetry. But 
this relation is also very vague at the moment.

In order to clarify the situation, we proceed in this section with some
differential geometry. 
We will try to avoid abstract formalisms and prefer calculations in explicit 
coordinate systems and gauges. This is less elegant than a 
coordinate-free formulation, but it leads to 
a more elementary and direct approach. Our constructions are helpful
for the physical interpretation, as will be explained in the next section
\ref{sec6}. In non-technical terms, our aim is to rewrite the Dirac operator
in a form which gives a better geometrical understanding of the matrices
$G^j(x)$,$B(x)$.

Before starting the mathematical analysis, we point out that we consider
the Dirac operator (apart from the wave functions of the physical particles)
as the only a-priori given object on the manifold. This means
that all additional structure (like the metric, curvature, classical
potentials, and field tensors) must be constructed from $G$.
It might seem unusual to take only the Dirac operator as the starting point of
a classical gauge theory. The motivation for this very restrictive procedure is
the following:
A physical theory is more convincing if it is developed from few objects, 
which are given from the very beginning and are considered as the 
``fundamental'' objects of the theory.
In relativistic quantum mechanics, the Dirac operator is needed in any case for
the formulation of the Dirac equation. The objects of classical field theory,
however, either enter directly in the Dirac operator or can be expressed in
terms of derivatives of potentials occurring in $G$.
We conclude that it is both desirable and might be possible to consider them
as derived objects, which are only constructed from the Dirac operator for
a convenient formulation of the classical interactions.

The Lorentzian metric (\ref{4_4}) induces the Levi-Civita connection
$\nabla$, which gives a parallel transport of vector and tensor 
fields along geodesics. The geometry of space-time can be described as 
usual with the Riemannian curvature tensor $R^i_{jkl}$, and we could thus 
formulate general relativity by writing down Einstein's field equations.
This procedure only tells about part of the geometry, however.
We did not at all use the matrix $B(x)$ and did not take into account the
$U(2,2)$ gauge symmetry.
If we want to understand all the degrees of freedom of the Dirac operator 
geometrically, it is a better idea to study the parallel transport of spinors
(instead of tensor fields). In the infinitesimal version, this parallel transport
is given by a $U(2,2)$ gauge covariant derivative $D$ on the wave 
functions, which we call {\em{spin derivative}}. It is characterized by 
the condition that the derivative $D_j \Psi$ of a wave function shall again 
behave under gauge transformations according to (\ref{4_30}).
The basic question is if the Dirac operator induces a spin derivative.

For the construction of the spin derivative, we will analyze gauge 
transformations explicitly. In a special gauge and coordinate system, $D$ 
shall have the representation
\begin{equation}
\label{5_1}
D_j \;=\; \frac{\partial}{\partial x^j} - i C_j(x)
\end{equation}
with suitable $(4 \times 4)$-matrices $C_j(x)$. Under a gauge transformation
(\ref{4_30}), it transforms like
\begin{equation}
\label{5_2}
D_j \;\rightarrow\; U D_j U^{-1} \;=\; \partial_j - i U C_j 
U^{-1} \:+\: U(\partial_j U^{-1}) \spc .
\end{equation}
Thus our aim is to find matrices $C_j(x)$ satisfying the transformation rule
\begin{equation}
	C_j \;\rightarrow\; U C_j U^{-1} \:+\: i U (\partial_j U^{-1}) \spc .
	\label{5_3}
\end{equation}
These matrices must be formed out of $G^j(x)$,$B(x)$, which behave 
under gauge transformations like
\begin{equation}
	G^j \;\rightarrow\; U G^j U^{-1} \;\;\;,\spc
	B \;\rightarrow\; UBU^{-1} \:+\: i UG^j (\partial_j U^{-1}) \spc .
	\label{5)4}
\end{equation}
It is far from being obvious how an explicit formula for the matrices
$C_j$ should look like.
But we can already say something about its general structure:
since only first derivatives of $U$ occur in (\ref{5_3}), we will only use 
$G^j$,$B$, and first derivatives of the Dirac matrices $\partial_k G^l$ for the
construction.
As a consequence, we need not care about second derivatives of $U$; they 
will not enter in our calculations. Constant gauge transformations (i.e.\ 
transformations $U$ with $\partial_j U=0$) are also irrelevant, because 
they only describe a common unitary transformation $. \rightarrow U \:.\: 
U^{-1}$ of all matrices. The point of interest is the first order term $i 
U (\partial_j U^{-1})$ in (\ref{5_3}).

We introduce some notation: In analogy to the bilinear and 
pseudoscalar covariants of the Dirac theory, we define the matrices
\[ \sigma^{jk}(x) \;=\; \frac{i}{2} \: [G^j,\: G^k] \;\;\;,\spc
\rho(x) \;=\; \frac{i}{4!} \: \epsilon_{jklm} \:G^j \:G^k\: G^l\: G^m \spc . \]
In a local reference frame (\ref{4_21}), they coincide with the usual 
matrices $\sigma^{jk}$,$\gamma^5$ in the Dirac representation
(we use the notation $\rho$ instead of $\gamma^5$, because the
``tensor index'' $^5$ might be confusing on a manifold).
A $(4 \times 4)$-matrix is called {\em{even}} or {\em{odd}} if it 
commutes resp.\ anti-commutes with $\rho$. The matrices
\begin{equation}
	G^j\;\;\;,\;\;\;\; \rho G^j \;\;\;,\;\;\;\; \1 \;\;\;,\;\;\;\; i \rho
	\;\;\;,\;\;\;\; \sigma^{jk}
	\label{5_10}
\end{equation}
form a basis of the 16-dimensional (real) vector space 
of self-adjoint matrices (with respect to $\Sl .|. \Sr$).
$G^j$, $\rho G^j$ are odd; $\1$, $i\rho$, and $\sigma^{jk}$
are even.

Since we must expect the matrices $C_j$ to be complicated expressions in
$G^j$ and $B$, it is useful to first consider special gauges where these
expressions should have a simple form. For this purpose, we study the term
$\nabla_k G^j = \partial_k G^j + \Gamma^j_{kl} \:G^l$. Under coordinate 
transformations, it behaves like a tensor. Under gauge transformations, 
however, first derivatives of $U$ occur,
\begin{eqnarray}
	\nabla_k G^j \;\rightarrow \nabla_k (U G^j U^{-1}) & = &
	U (\nabla_k G^j) U^{-1} \:+\: (\partial_k U) G^j U^{-1} \:+\:
	U G^j (\partial_k U^{-1}) \nonumber \\
& = & U (\nabla_k G^j) U^{-1} \:-\: \left[ U (\partial_k U^{-1}),\:
	 U G^j U^{-1} \right] \spc .
	\label{5_5a}
\end{eqnarray}
We can use the second summand in (\ref{5_5a}) for a partial gauge fixing:
\begin{Lemma}
\label{lemma1}
For any space-time point $p \in M$ there is a gauge such that
\begin{equation}
	\nabla_k G^j(p) \;=\; 0 \spc .
	\label{5_5}
\end{equation}
\end{Lemma}
{\Proof}
Notice that the matrix $\partial_j \rho$ is odd, because
\begin{equation}
\label{5_6}
0 \;=\; \partial_j \1 \;=\; \partial_j (\rho \rho) \;=\;
(\partial_j \rho) \rho + \rho (\partial_j \rho) \spc .
\end{equation}
We start with an arbitrary gauge and construct the desired gauge with two 
subsequent gauge transformations:
\begin{enumerate}
	\item  According to (\ref{5_6}), the matrix $i \rho (\partial_j \rho)$ is 
	self-adjoint. We can thus perform a gauge 
	transformation $U$ with $U(p)=\1$, $\partial_j U(p)=\frac{1}{2} \rho 
	(\partial_j \rho)$. The matrix $\partial_j \rho(p)$ vanishes in the
	new gauge, since
\[ \partial_j \rho_{|p} \rightarrow \partial_j (U \rho U^{-1})_{|p} \;=\; 
\partial_j \rho_{|p} + \frac{1}{2} \left[ \rho (\partial_j \rho), \: \rho 
\right]_{|p} \;=\; \partial_j \rho_{|p} - \rho^2 (\partial_j 
\rho)_{|p} \;=\; 0 \spc . \]
    By differentiating the relation $\{ \rho, G^j \}=0$, we conclude that
    the matrix $\nabla_k G^j_{|p}$ is odd. We can thus represent it in the form
\begin{equation}
	\nabla_k G^j_{|p} \;=\; \Lambda^j_{km} \:G^m_{|p} \:+\: 
\Theta^j_{km} \: \rho G^m
	\label{5_7}
\end{equation}
    with suitable coefficients $\Lambda^j_{km}$, $\Theta^j_{km}$.
    
    This representation can be further simplified: According to Ricci's 
    Lemma, we have at $p$
\begin{eqnarray}
	0 & = & 2 \nabla_n g^{jk} \;=\; \{ \nabla_n G^j,\:G^k\} \:+\: \{ G^j,\: 
	\nabla_n G^k\} \nonumber \\
	 & = & 2 \Lambda^j_{nm} \: g^{mk} \:-\: \Theta^j_{nm} \: 2i\rho \sigma^{mk}
	 \:+\: 2 \Lambda^k_{nm} \: g^{mj} \:-\: \Theta^k_{nm} \: 2i\rho \sigma^{mj}
	 \label{5_9} \;\;\; ,
\end{eqnarray}
    and thus
\begin{equation}
	\Lambda^j_{nm} \:g^{mk}_{|p} \;=\; -\Lambda^k_{nm} \: g^{mj}_{|p} \spc .
	\label{5_8}
\end{equation}  
    In the case $j=k \neq m$, (\ref{5_9}) yields that $\Theta^j_{nm}=0$.
    For $j \neq k$, we obtain $\Theta^j_{nj} \:\sigma^{jk} + 
    \Theta^k_{nk} \:\sigma^{kj} = 0$ and thus $\Theta^j_{nj} = \Theta^k_{nk}$
    ($j$ and $k$ denote fixed indices, no summation is performed).
    We conclude that there are coefficients $\Theta_k$ with
\begin{equation}
	\Theta^j_{km} \;=\; \Theta_k \: \delta^j_m \spc .
	\label{5_9a}
\end{equation}
	\item  We perform a gauge transformation $U$ with $U(p)=\1$ and
\[ \partial_k U \;=\; -\frac{1}{2} \:\Theta_k \: \rho \:-\: \frac{i}{4} \: 
\Lambda^m_{kn} \: g^{nl} \: \sigma_{ml} \spc . \]
    Using the representation (\ref{5_7}) together with 
    (\ref{5_8}),(\ref{5_9a}),
    the matrix $\nabla_k G^j$ transforms like
\begin{eqnarray*}
	\spc \nabla_k G^j & \rightarrow & \nabla_k G^j \:+\: [\partial_k U,\: G^j]  \\
	 &  & =\; \Lambda^j_{km} \:G^m \:+\: \Theta_k \: \rho G^j
	 \:-\: \Theta_k \: \rho G^j \:-\: 
	 \frac{i}{4}\: \Lambda^m_{kn} \:g^{nl} \: [\sigma_{ml},\: G^j] \\
	 &  & =\; \Lambda^j_{km} \:G^m \:-\: \frac{i}{4}\: \Lambda^m_{kn} 
	 \:g^{nl} \:2i \: (G_m \: \delta^j_l - G_l \:\delta^j_m) \\
	 &  & =\; \Lambda^j_{km} \:G^m \:+\: \frac{1}{2}\: \Lambda^m_{kn} \: 
	 g^{nj} \: G_m \:-\: \frac{1}{2}\: \Lambda^j_{km} \: G^m \;=\; 0 \spc .
	 \spc \;\;\; \FBox
\end{eqnarray*}
\end{enumerate}
In general, the condition $\nabla_k G^j$ can only be satisfied ``locally'' 
in one point $p \in M$. We call a gauge satisfying (\ref{5_5}) a 
{\em{normal gauge}} around $p$.

Now we look to which extent the gauge is fixed by Lemma \ref{lemma1}.
According to (\ref{5_5a}), a transformation between normal gauges must 
satisfy the condition $[U (\partial_j U^{-1}),\: UG^jU^{-1}]_{|p}=0$.
As a consequence, the matrix $iU (\partial_j U^{-1})_{|p}$ must be a multiple 
of the identity, as is verified in the basis (\ref{5_10}) using the
commutation relations between the Dirac matrices.
Since constant gauge transformations and higher than first order derivatives of
$U$ are irrelevant, we can assume that $U(x)$ itself is a multiple of the identity.
In other words, we can restrict ourselves to $U(1)$ gauge
transformations. This is very helpful because in this special case there is a
simple expression showing the transformation law (\ref{5_3}), namely  
\begin{eqnarray*}
	\frac{1}{4}\: {\mbox{Re }} \Tr (G_j \:B) \: \1  & \rightarrow & 
	\frac{1}{4}\: {\mbox{Re }} \Tr (G_j \:B) \: \1 \:+\:
	\frac{1}{4}\: {\mbox{Re }} \Tr \left( G_j \:G^k \: iU(\partial_j U^{-1}) \:
	  \1 \right) \\
	 &  & =\; \frac{1}{4}\: {\mbox{Re }} \Tr (G_j \:B) \: \1 \:+\:
	 i U (\partial_j U^{-1}) \spc .
\end{eqnarray*}
We can identify this expression with $C_j$ and use (\ref{5_1}) as the definition
for a spin derivative:
\begin{Def}
We define the {\bf{canonical spin derivative}} $D$ by the condition that 
it has in normal gauges around $p$ the form
\begin{equation}
	D_j(p) \;=\; \frac{\partial}{\partial x^j} \:-\: \frac{i}{4} \:
	{\mbox{Re }} \Tr (G_j B)_{|p} \: \1 \spc .
	\label{5_11}
\end{equation}
\end{Def}
In general gauges, the canonical spin derivative can be written as
\begin{equation}
	D_j \;=\; \frac{\partial}{\partial x^j} \:-\: i E_j \:-\: \frac{i}{4} \:
	{\mbox{Re }} \Tr (G_j B) \: \1
	\label{5_12}
\end{equation}
with additional matrices $E_j(x)$. These matrices are characterized by the
condition that $E_j(p)$ vanishes in normal gauges around $p$ and behaves according
to (\ref{5_3}) under $SU(2,2)$ gauge transformations.
They are given by the formula
\begin{equation}
	E_j \;=\; \frac{i}{2}\: \rho (\partial_j \rho) \:-\: \frac{i}{16}\: \Tr 
	(G^m \:\nabla_j G^n) \: G_m G_n \:+\: \frac{i}{8}\: \Tr (\rho G_j \:
	\nabla_m G^m) \:\rho \spc ,
	\label{5_12a}
\end{equation}
as can be verified by a straightforward calculation. Unfortunately, this
expression is rather complicated. We will in the following always avoid
working with $E_j$ explicitly.

The canonical spin derivative has some nice properties. We first collect them
in a theorem and discuss them afterwards.
\begin{Thm}
\label{thm_2}
The canonical spin derivative satisfies the equation
\begin{equation}
	[D_k, G^j] \:+\: \Gamma^j_{kl} \: G^l \;=\; 0 \spc .
	\label{5_13}
\end{equation}
It is compatible with the spin scalar product,
\begin{equation}
	\partial_j \:\Sl \Psi \:|\: \Phi \Sr \;=\; \Sl D_j \Psi \:|\: \Phi \Sr
	\:+\: \Sl \Psi \:|\: D_j \Phi \Sr \spc .
	\label{5_17}
\end{equation}
The operator $i G^j D_j$ is a well-defined Hermitian operator on $H$.
Furthermore,
\begin{equation}
	-i \nabla_j \:\Sl G^j \Psi \:|\: \Phi \Sr \;=\; \Sl i G^j D_j \Psi \:|\: \Phi \Sr
	\:-\: \Sl \Psi \:|\: i G^j D_j \Phi \Sr \spc .
	\label{5_15}
\end{equation}
\end{Thm}
{\Proof}
The left side of (\ref{5_13}) behaves under gauge transformations according to
the adjoint representation $. \rightarrow U \:.\: U^{-1}$ of the gauge group.
Thus it suffices to check (\ref{5_13}) in a normal gauge, where
\[ [D_k,G^j] + \Gamma^j_{kl} \: G^l \;=\; \nabla_k G^j \:-\: \frac{i}{4}\:
   {\mbox{Re }} \Tr (G_j B) \: [\1,G^j] \;=\; 0 \spc . \]
   
For the canonical spin derivative, the matrix $C_j$ in (\ref{5_1}) is 
self-adjoint, as can be verified with (\ref{5_12}),(\ref{5_12a}).
This immediately implies (\ref{5_17}).

According to its behavior under coordinate and gauge transformations, we can
view $i G^j D_j$ as an operator on $H$. Relation (\ref{5_17}) yields the equation
\[ \nabla_j \Sl G^j \Psi \:|\: \Phi \Sr \;=\; \Sl D_j G^j \Psi \:|\: \Phi 
	\Sr \:+\: \Sl G^j \Psi \:|\: D_j \Phi \Sr \:+\: \Gamma^j_{jk} \:
	\Sl G^k \Psi \:|\: \Phi \Sr \spc . \]
Applying (\ref{5_13}) and the self-adjointness of the Dirac matrices, we
obtain (\ref{5_15}).
If we integrate this equation with respect to the invariant measure
$d\mu = \sqrt{|g|}\: d^4x$, the left side vanishes with Gauss' theorem.
On the right side, we substitute (\ref{4_00}) and conclude that the operator
$i G^j D_j$ is Hermitian.
\QED
Generally speaking, this theorem shows that the connections $\nabla$,$D$,
the Dirac matrices $G^j$, and the scalar products $g_{jk}$,$\Sl .|. \Sr$ can
be consistently used in combination with each other, if all expressions are
written in a coordinate and gauge invariant form (by appropriately using
$\nabla_j$, $D_j$, and $\Gamma^j_{kl}$). We can leave out
the derivatives of $G^j$ and may use the product rule inside the spin scalar product.
Relation (\ref{5_17}) formally corresponds to Ricci's Lemma in Riemannian
geometry if we replace the metric by the spin scalar product and $\nabla$ by $D$.

Since the operators $G$ and $i G^j D_j$ are both Hermitian and coincide up to
zero order contributions, we can represent the Dirac operator in the form
\begin{equation}
	G \;=\; i G^j D_j + H
	\label{5_50}
\end{equation}
with a self-adjoint matrix $H(x)$.
Under gauge transformations, $H$ simply behaves according to the adjoint
representation. This is an advantage over the formula (\ref{4_6}),
where the transformation of $B$ involves derivatives of $U$.

Equation (\ref{5_15}) implies a general version of current conservation:
Assume that $\Psi$ and $\Phi$ are eigenvectors of $G$ with real eigenvalue 
$m$. Then the vector field
$\Sl G^j \Psi \:|\: \Phi \Sr$ is divergence-free,
\[ -i \nabla_j \Sl G^j \Psi \:|\: \Phi \Sr \;=\; \Sl (G-H) \Psi \:|\: \Phi \Sr
   \:-\: \Sl \Psi \:|\: (G-H) \Phi \Sr \;=\; 0 \spc . \]
It is a natural generalization of the electromagnetic current 
$\overline{\Psi} \gamma^j \Psi$ of the Dirac theory.
According to Gau{\ss}' law, the current conservation implies that the 
scalar product (\ref{SP}) is independent of the choice of the 
hypersurface ${\cal{H}}$. The integrand of the scalar product $(\Psi 
\:|\: \Psi)$ has the interpretation as the probability density of the 
particle.

The Christoffel symbol in (\ref{5_13}) indicates that the spin derivative in
some way includes the Levi-Civita connection $\nabla$. The following bundle
construction makes this relation more precise:
We denote the sections of a vector bundle $N$ over $M$ by $\Gamma(N)$
(e.g.\ $\Gamma(TM)$ are the vector fields on $M$).
The Levi-Civita connection is a connection on the tangent bundle $TM$,
\[ \nabla \;:\; T_pM \times \Gamma(TM) \rightarrow T_pM \;:\;
   (X, Y) \rightarrow \nabla_X Y \spc . \]
We can view the wave functions as sections of a vector bundle $SM$ with 
fibre $(\C^4,\: \Sl .|. \Sr)$. In order to describe the $U(2,2)$ gauge symmetry, we 
make this {\em{spin bundle}} to a principal bundle with structure group $U(2,2)$,
which acts on the fibre in the fundamental representation $. \rightarrow
U.$. The spin derivative is a connection on the spin bundle,
\[ D \;:\; T_pM \times \Gamma(SM) \rightarrow S_pM \;:\; (X, \Psi) 
\rightarrow D_X \Psi \spc . \]
We define $L_pM$ as the vector space of self-adjoint transformations of
$S_pM$. In a special gauge, $L_pM$ can be represented as $(4 \times 4)$-matrices; 
(\ref{5_10}) gives an explicit basis. The gauge transformations act on the
matrices according to the adjoint representation $. \rightarrow 
U_{|p} \:.\: U^{-1}_{|p}$. We denote the corresponding principal bundle by
$LM$. The spin derivative induces a connection on $LM$ by
\begin{equation}
	D \;:\; T_pM \times \Gamma(LM) \rightarrow L_pM \;:\; (X, A) \rightarrow
	[D_X, A] \spc .
	\label{5_40}
\end{equation}
We can view the tangent bundle as a subspace of $LM$, because there 
are canonical embeddings and projections
\begin{eqnarray*}
	\iota_p & : & T_pM \hookrightarrow L_pM \;:\; X \rightarrow X^i \: G_i  \\
	\pi_p & : & L_pM \rightarrow T_pM \;:\; A \rightarrow \frac{1}{4} \:
	\Tr (G^j A) \: \frac{\partial}{\partial x^j}
\end{eqnarray*}
with $\pi \iota = {\mbox{id}}$. By restricting the connection (\ref{5_40}) to the
tangent bundle, we can form a connection $\tilde{\nabla}$ on $TM$,
\[ \tilde{\nabla} \;:\; T_pM \times {\cal S}(TM) \rightarrow T_pM \;:\;
   (X, Y) \rightarrow \pi \:[D_X,\: \iota Y] \spc . \]
Equation (\ref{5_13}) states that this connection coincides with the 
Levi-Civita connection, since
\begin{eqnarray}
	[D_X,\: \iota Y] & = & X^j \: [D_j,\: G^k \:Y_k] \;=\; X^j \left(
	(\partial_j Y_k) \:G^k \:-\: \Gamma^k_{jl} \:Y_k \:G^l \right) \nonumber \\
	 & = & X^j \: (\nabla_j Y_k) \: G^k \;=\; \iota \:\nabla_X Y
     \label{5_41}
\end{eqnarray}
and thus
\[ \tilde{\nabla}_X Y \;=\; \pi \:[D_X,\: \iota Y] \;=\; \pi \iota \:
	\nabla_X Y \;=\; \nabla_X Y \spc . \]
In this way, the Levi-Civita connection is recovered as a suitable 
restriction of the spin connection. The geometry of the spin bundle
generalizes the Lorentzian geometry of the manifold.

It remains to construct the geometrical invariants of the
spin connection. On an elementary level, one can look for expressions 
in $D_k$,$G^j$ which transform both like a tensor and according to a local
representation of the gauge group. This leads to the definition of the 2-forms
\begin{eqnarray*}
	T_{jk} & = & \frac{i}{2} \left( [D_j, G_k] - [D_k, G_j] \right)  \\
	R_{jk} & = & \frac{i}{2} \: [D_j, D_k] \spc ,
\end{eqnarray*}
which are called {\em{torsion}} and {\em{curvature}}.
\begin{Thm}
\label{thm2}
The canonical spin connection is torsion-free. Curvature has the form
\begin{equation}
	R_{jk} \;=\; \frac{1}{8} \: R_{mnjk} \:\sigma^{mn} \:+\: \frac{1}{2} \: 
    (\partial_j a_k - \partial_k a_j) \: \1
	\label{5_21}
\end{equation}
with the Riemannian curvature tensor $R_{mnjk}$ and the potential
\begin{equation}
a_j \;=\; \frac{1}{4}\: {\mbox{Re }} \Tr (G_j B) \spc .
\label{5_49}
\end{equation}
\end{Thm}
{\Proof}
Relation (\ref{5_13}) yields
\[ [D_j,\: G_k] \;=\; [D_j, \:g_{kl} \: G^l] \;=\; (\partial_j g_{kl}) \: G^l
   \:-\: g_{kl} \: \Gamma^l_{jm} \: G^m \;=\; \Gamma^m_{jk} \: G_m \]
and thus, using that the Levi-Civita connection is torsion-free,
\[ T_{jk} \;=\; \frac{i}{2} \: (\Gamma^m_{jk} - \Gamma^m_{kj}) \:
G_m \;=\; 0 \spc . \]

We iterate formula (\ref{5_41}) and express the Riemannian curvature 
tensor in terms of the curvature of the spin connection,
\begin{eqnarray*}
    G_i \: R^i_{jkl} \:Z^l &=& \iota \: (\nabla_j \nabla_k Z - \nabla_k
    \nabla_j Z) \\
    &=& [D_j,\: [D_k,\: \iota Z ]] - [D_k,\: [D_j,\: \iota Z ]] \\
	 & = & [[D_j, \: D_k],\: \iota Z] \;=\; -2i \: [R_{jk},\: \iota Z] \spc .
\end{eqnarray*}
This equation determines curvature up to a multiple of the identity 
matrix,
\[ R_{jk}(x) \;=\; \frac{1}{8} \: R_{mnjk} \:\sigma^{mn} \:+\:
   \lambda_{jk} \1 \spc . \]
Thus it remains to calculate the trace of curvature,
\[ \frac{1}{4}\: \Tr (R_{jk}) \;=\; \frac{1}{8}\:
     \Tr (\partial_j C_k - \partial_k C_j)  \;=\; \frac{1}{2} \:
     (\partial_j a_k - \partial_k a_j) \spc , \]
where we used (\ref{5_12}) and the fact that the matrices $E_j$ are 
trace-free.
\QED
We remark that the canonical spin derivative is not the only spin derivative
which can be constructed from the Dirac operator.
A general spin derivative $\tilde{D}$ differs from the canonical spin derivative
by matrices $F_j=\tilde{D}_j - D_j$ which transform according to the adjoint
representation of the gauge group.
It is no loss of generality to work with the canonical spin derivative because
the matrices $F_j$ can also be dealt as additional potentials. Because of
the useful properties of Theorem \ref{thm2}, it is most convenient
to use the canonical spin derivative.

\section{Interpretation, the Physical Dirac Operator}
\label{sec6}
For the physical interpretation of the previous constructions,
we return to the discussion of the moving observer on page \pageref{obs}.
We saw that the coordinate system and gauge satisfying (\ref{4_21}) give a
local reference frame for the observer.
The gravitational field can be compensated locally by choosing a normal
coordinate system. According to Lemma \ref{lemma1}, we can further compensate
the dynamics of the Dirac matrices: in a normal gauge even the derivatives
of the Dirac matrices vanish, $\partial_k G^j(p)=\nabla_k G^j(p)=0$.
This means that the above
mentioned ``spin mixing force'' can also be compensated locally by choosing
an appropriate coordinate system and gauge. We cannot do even better because
second order derivatives of the metric and the Dirac matrices involve the
Riemannian curvature tensor. We call a coordinate system
and gauge satisfying the conditions $G^j(p)=\gamma^j$, $\partial_k G^j(p)=0$ a
{\em{normal reference frame}} around $p$.

We come to the discussion of the matrix $B(x)$, which was not at all specified
in Def.\ \ref{def1}. The strongest local condition is to impose that $B(p)$
vanishes in a suitable normal reference frame around $p$.
This condition turns out to be of physical interest,
and we use it as the definition for the physical Dirac operator:
\begin{Def}
\label{def_pdo}
A Dirac operator $G$ is called {\bf{physical Dirac operator}} if for any $p \in M$
there is a normal reference frame around $p$ such that $B(p)=0$.
\end{Def}
In other words, the physical Dirac operator is characterized by the condition
that it locally coincides with the free Dirac operator, which means more
precisely that there is a coordinate system and a gauge with $G^j(p)=\gamma^j$,
$\partial_k G^j(p)=0$ and $B(p)=0$.

We again consider the physical Dirac operator as the only a-priori given object on
the manifold and construct the metric and the canonical spin derivative from $G$.
In a normal reference frame with $G=i G^j \partial_j$, the canonical spin
derivative (\ref{5_11}) coincides with the partial derivatives, $D_j = \partial_j$.
Thus the physical Dirac operator has the representation
\begin{equation}
G \;=\; i G^j D_j \spc ,
\label{50}
\end{equation}
which is also valid in a general gauge and coordinate system.
In comparison to the representation (\ref{4_6}), the degrees of freedom
of the matrix $B(x)$ now occur in the canonical spin derivative.
This has the advantage that they can be immediately identified with the
$U(2,2)$ gauge potentials in (\ref{5_12}).
The last summand in (\ref{5_12}) describes the $U(1)$ phase transformations
of electrodynamics. We define the electromagnetic potential by
\[ A_j \;=\; \frac{1}{4e} \: {\mbox{Re }} \Tr (G_j \:B) \]
and rewrite the canonical spin derivative in the more familiar form
\[ D_j \;=\; \frac{\partial}{\partial x^j} \:-\: i E_j \:-\: ie A_j 
\spc . \]
The matrices $E_j$ take into account the gravitational field.
According to Theorem \ref{thm2}, the curvature of the spin connection is formed
of the Riemannian curvature tensor and the
electromagnetic field tensor $F_{jk} = \partial_j A_k - \partial_k A_j$.
We can write down the classical action in terms of these tensor fields.
The classical variational principle yields a unified description of general
relativity and electromagnetism as a $U(2,2)$ gauge symmetry.

The definition of the physical Dirac operator can also be understood in more
technical terms with the canonical spin derivative:
The Dirac operator of Def.\ \ref{def1} has the representation
(\ref{5_50}) with a self-adjoint matrix $H(x)$. According to the construction
of the spin derivative, the matrices $G^j H$ are trace-free. Thus $H(x)$ is
characterized by twelve degrees of freedom, which correspond to additional 
potentials (more precisely, 1 scalar, 1 pseudoscalar, 4 pseudovector, and 6
bilinear potentials). These potentials are not gauge potentials. They do not seem
to occur in nature. Therefore it is a reasonable physical condition to assume
that $H$ vanishes. This leads to the representation (\ref{50}) of the
physical Dirac operator.

Our definition of the Dirac operator introduces the gauge potentials and
in this way replaces the usual minimal coupling procedure.
We explain for clarity why simple minimal coupling does not make sense 
in the context of our $U(2,2)$ symmetry: Assume that we had (instead of constructing 
the spin derivative from the Dirac operator) introduced a $U(2,2)$ gauge
covariant derivative $D$ of the form (\ref{5_1}) with gauge potentials 
$C_j$. According to minimal coupling, we must replace the
partial derivatives in the free Dirac operator by gauge covariant 
derivatives, which gives the operator $i \gamma^j D_j$.
This operator does not make sense, however.
First of all, it is not Hermitian (with respect to $\bra .|. \ket$).
Furthermore, the current $\Sl \gamma^j \Psi \:|\: \Psi \Sr$ is in 
general not divergence-free, so that the probabilistic interpretation 
of the Dirac wave function breaks down. The basic reason why minimal coupling
does not work is that the $U(2,2)$ gauge potentials $C_j$ do in general not
commute with the Dirac matrices $\gamma^j$. In order to bypass these 
problems, we must replace the $\gamma$-matrices by dynamical Dirac
matrices $G^j(x)$. The two replacements
$\partial_j \rightarrow D_j$ and $\gamma^j \rightarrow G^j$ must be
coordinated in such a way that the Dirac operator $i G^j D_j$ is 
Hermitian and allows a reasonable definition of a conserved current.
This is accomplished by our definition of the Dirac operator and
the subsequent construction of the spin derivative.

Our description of gravitation differs considerably from the usual
formulation of general relativity as a gauge theory (see e.g.\ \cite{GS}).
As one of the major differences, we avoid the principal bundle of orthonormal
frames (with the Lorentz group as structure group); instead we are working with
the spin bundle and local $U(2,2)$ transformations.
This replacement is possible because the subgroup of $U(2,2)$ generated by
the bilinear covariants $\sigma^{ij}$ is locally isomorphic to the Lorentz group.
In contrast to the connection on the bundle of orthonormal frames, the canonical
spin connection is torsion-free. Remember that the $U(2,2)$ spin bundle arose as
a consequence of our measurement principle for space and time. Thus, as
an advantage of our procedure, the usual bundle constructions for the definition
of gauge fields and of spinors in curved space-time are no longer necessary.

\section{Conclusion}
We saw that the adaptation of the ideas in \cite{F1} to the relativistic context
yields a local $U(2,2)$ gauge symmetry of the Dirac equation.
In order to describe the physical interactions with this gauge symmetry,
it is necessary to consider the Dirac operator as the basic object on the manifold.
The gauge potentials are implicitly contained in
the Dirac operator. By constructing the spin derivative, they are recovered as
describing the electromagnetic and
gravitational field. In this way, we conclude that the local $U(2,2)$ symmetry
in relativistic quantum mechanics makes  physical sense.
The concept of measurability of space-time gives a fundamental explanation
for this gauge symmetry.

Our description has the advantage that both the Dirac theory and classical field
theory are developed from few a-priori given objects:
The fermionic particles correspond to vectors of the indefinite scalar product
space $(H, \bra .|. \ket)$. Space-time is described by the spectral measure
$dE_x$ on the manifold $M$.
The Dirac operator gives the gauge potentials and determines the interaction
between the fermions and the gauge fields.
This description is conceptually simple. It is the starting point for further
constructions which finally lead to the ``Principle of the Fermionic Projector''
as introduced in \cite{F2}.

\end{document}